\documentclass[aps,prl,reprint,superscriptaddress]{revtex4-1}

\usepackage{graphicx}
\usepackage{epsfig}
\usepackage{placeins}
\usepackage{amsmath}
\usepackage{nicefrac}
\usepackage{lineno}

\begin{document}

\title{Current-controlled Spin Precession of Quasi-Stationary Electrons in a Cubic Spin-Orbit Field}
\author{P.~Altmann}
\affiliation{IBM Research--Zurich, S\"aumerstrasse 4, 8803 R\"uschlikon, Switzerland}

\author{F.~G.~G.~Hernandez}
\affiliation{Instituto de Fisica, Universidade de S\~ao Paulo, S\~ao Paulo 05508-090, S\~ao Paulo, Brazil}

\author{G.~J.~Ferreira}
\affiliation{Instituto de Fisica, Universidade Federal de Uberl\^andia, Uberl\^andia 38400-902, Minas Gerais, Brazil}

\author{M.~Kohda}
\affiliation{Department of Materials Science, Tohoku University, 6-6-02 Aramaki-Aza Aoba, Aoba-ku, Sendai 980-8579, Japan}

\author{C.~Reichl}
\affiliation{Solid State Physics Laboratory, ETH Zurich, 8093 Zurich, Switzerland}

\author{W.~Wegscheider}
\affiliation{Solid State Physics Laboratory, ETH Zurich, 8093 Zurich, Switzerland}

\author{G.~Salis}
\email{gsa@zurich.ibm.com}
\affiliation{IBM Research--Zurich, S\"aumerstrasse 4, 8803 R\"uschlikon, Switzerland}

\begin{abstract}
Space- and time-resolved measurements of spin drift and diffusion are performed on a GaAs-hosted two-dimensional electron gas. For spins where forward drift is compensated by backward diffusion, we find a precession frequency in absence of an external magnetic field. The frequency depends linearly on the drift velocity and is explained by the cubic Dresselhaus spin-orbit interaction, for which drift leads to a spin precession angle twice that of spins that diffuse the same distance.
\end{abstract}

\maketitle

Drift and diffusion of charge carriers in semiconductor nanostructures are the foundation of information technology. The spin of the electron is being investigated as an additional or complementary degree of freedom that can enhance the functionality of electronic devices and circuits~\cite{ZuticFabianDasSarma, Awschalom2007, Datta2010}. 
In the presence of spin-orbit interaction (SOI), the spins of moving electrons precess about effective magnetic fields that depend on the electron momentum vector, $\mathbf{k}$~\cite{Winkler}. 
In a two-dimensional electron gas (2DEG), this precession has been proposed as a gate-tunable switching mechanism~\cite{Datta1990, Chuang2014}. 
Spin diffusion and spin drift have been studied using optical~\cite{Cameron1996, Kikkawa1999, Crooker2005, Wunderlich2010, Voelkl2011} and electrical techniques~\cite{Dresselhaus1992, Lou2007}. 
A local spin polarization expands diffusively into a spin mode with a spatial polarization pattern that is characteristic of the strength and symmetry of the SOI~\cite{Stanescu_2007}. An additional drift induced by an electric field does not modify the spatial precession period in the case of linear SOI~\cite{Hruska2006, Cheng2007, Yang2010, Yang2012}. This is because spins that travel a certain distance and direction precess on average by the same angle, irrespective of how the travel is distributed between diffusion and drift. Therefore, no spin precession occurs for quasi-stationary electrons, i.e. for electrons where drift is compensated by diffusion.

\begin{figure}[ht!]
\includegraphics[width=\columnwidth]{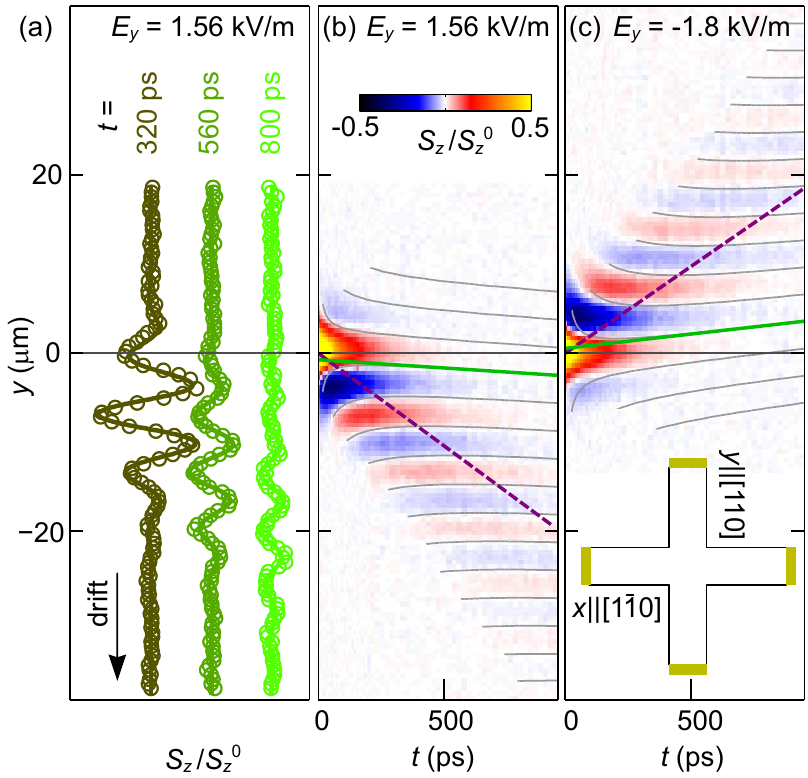}
\caption{Measurement of drifting spins after a local spin excitation at time $t=0$. (a) Measured spin polarization $S_z$ vs. $y$ for different $t$ at an electric field $E_y = 1.56$~kV/m. The data is offset according to $t$ and normalized to the maximum spin polarization, $S_z^0$. Circles are experimental data and solid lines are fits with Eq.\ (\ref{eq:fit}). (b) Colorscale plot of $S_z(y,t)$ for $E_y = 1.56$~kV/m. The violet dashed line marks the center of the spin packet. The gray solid lines are contour lines of a global fit as explained in the text. The solid green line indicates the slope of the lines of equal spin phase. It is tilted because spin precession from drift is different from that from diffusion owing to cubic SOI. (c) Colorscale plot of $S_z(y,t)$ for $E_y = -1.8$~kV/m, where the slope of the green line is reversed. Inset: schematic layout of the cross-shaped mesa structure. Four ohmic contacts allow the application of electric fields along the $y || [110]$ and the $x || [1\bar{1}0]$ direction.
\label{fig2} }
\end{figure}

In this letter, we experimentally observe such unexpected drift-induced spin precession of stationary electron spins in the absence of an external magnetic field. Using an optical pump-probe technique, we investigate the spatiotemporal dynamics of locally excited spin polarization in an n-doped GaAs quantum well. Spin polarization probed at a fixed position is found to precess with a finite frequency, $\omega$. This is identified as a consequence of cubic SOI, which affects spin drift and spin diffusion differently. A simple model predicts that drifting spins precess twice as much as spins that diffuse the same distance. This difference leads to a dependence $\omega \propto \beta_3 v_\mathrm{dr}$, where $\beta_3$ is the cubic SOI coefficient and $v_\mathrm{dr}$ the drift velocity. We demonstrate quantitative agreement between model and experiment, and extract a $\beta_3$ in agreement with literature values. Monte-Carlo simulations confirm the validity of the model and pinpoint deviations that occur when the drift-induced SOI field is small compared with that from diffusion into a perpendicular direction. This finding highlights the role of nonlinear SOI in spin transport and is relevant for spintronics applications.

\begin{figure}
\includegraphics[width=\columnwidth]{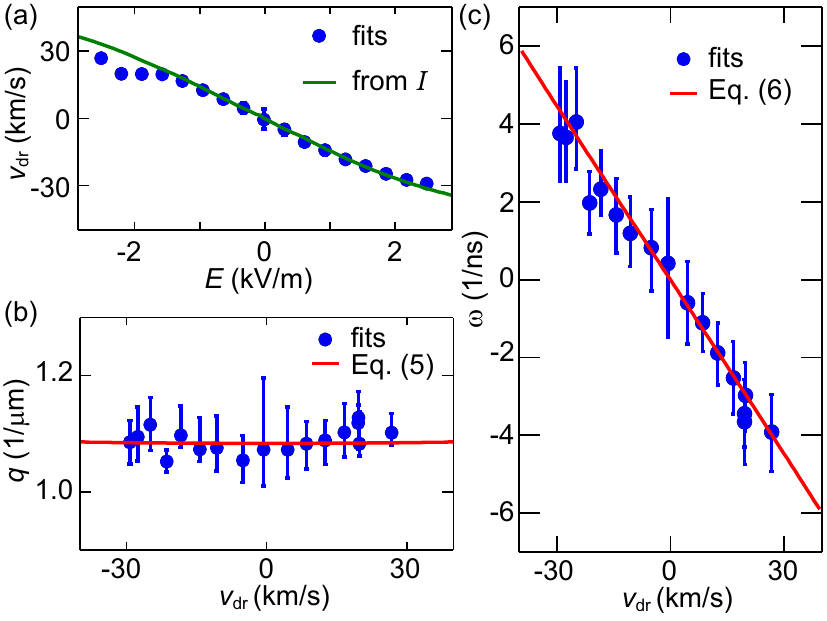}
\caption{Fit results. (a) Drift velocity $v_\mathrm{dr}$ plotted against the applied electric field. Dots are the fit values obtained from the measured $S_z (y,t)$. The solid line is the drift velocity calculated from the measured current $I$ via $v_\mathrm{dr} = I /(e n_\mathrm{s} w)$. (b) Values for the spatial wavenumber, $q$. Dots are the fit values and the red line is the model of Eq.\ (\ref{eq:q}) with $\alpha +\beta^* = 6.2 \times 10^{-13}$~eVm. (c) Values for the precession frequency, $\omega$. Dots are fit values and the red line is the model of Eq.\ (\ref{eq:omg}) with $\beta_3 = 8.5 \times 10^{-14}$~eVm. Confidence intervals in all plots are defined as a 5\% increase of the fit error. 
\label{fig3} }
\end{figure}

The sample consists of a 12-nm-thick GaAs quantum well in which the SOI is tuned close to the persistent spin helix (PSH) symmetry~\cite{Schliemann2003, Bernevig2006}. There, the effective magnetic field from linear SOI is strongly anisotropic, such that diffusing spins exhibit a strong spatial precession along the $y = [110]$ direction and no precession along $x = [1\bar{1}0]$ \cite{Walser2012}. The 2DEG has a sheet density of $n_\mathrm{s} = 5 \times 10^{15}~\mathrm{m}^{-2}$ with one occupied subband and a mobility of $22~\mathrm{m^2 (Vs)^{-1}}$, as determined by a Van-der-Pauw measurement at 4~K after illumination. Further details on the sample structure are given in Ref.~[\onlinecite{Walser2012}]. A cross-shaped mesa structure [cf. inset in Fig.~\ref{fig2}(c)] with a width $w = 150~\mathrm{\mu m}$ was fabricated by photo lithography and wet-chemical etching. We applied an in-plane electric field $E_y$ to the 2DEG along $y$ via two ohmic contacts, which are 800~$\mu$m apart. Spins oriented along the $z$-axis were locally excited in the center of the mesa at time $t=0$ by an optical pump pulse. At varying time-delay, $t$, the transient spin polarization along the $z$-axis, $S_z (y, t)$, was measured using the pump-probe technique described in~\cite{Walser2012, Altmann2014, Altmann2015} with a spatial resolution of $< 2 \mathrm{\mu m}$. The time-averaged laser power of the pump (probe) beam was $150~\mathrm{\mu W}$ ($15~\mathrm{\mu W}$) at a repetition rate of 80~MHz. The sample temperature was 20~K. Figure \ref{fig2}(a) shows data for three different time delays, $t$, at $E_y = 1.56$~kV/m. The spatially precessing spins are well described by a cosine oscillation in a Gaussian envelope, which broadens with time because of diffusion. The center of the envelope shifts along $-y$ because the electrons drift in the applied electric field. 
Figs.\ \ref{fig2}(b) and \ref{fig2}(c) show colorscale plots of $S_z (y, t)$ for $E_y = 1.56$~kV/m and $E_y = -1.8$~kV/m, respectively. The motion of the center of the spin packet is marked by a violet dashed line. Remarkably, the position of constant spin precession phase shifts along $y$ in time, as indicated by the solid green lines. This corresponds to a finite temporal precession frequency $\omega$ for spins that stay at a constant position $y$. 
For a positive $E_y$ [Fig.\ \ref{fig2}(b)], the spin packet moves towards the negative $y$-axis, and the tilt $\partial y/\partial t$ of constant spin phases is negative. Both the drift direction and the tilt change their sign when the polarity of $E_y$ is reversed [Fig.\ \ref{fig2}(c)]. 

We model $S_z$ by multiplying the Gaussian envelope by $\cos (q y + \omega t)$ and a decay factor $\exp (-t/\tau)$:

\begin{align}
\begin{split}
S_z (y, t) =   \frac{A_0}{2 D_\mathrm{s} t} \exp  \left[\frac{-(y - v_\mathrm{dr} t)^2}{ 4 D_\mathrm{s} t} \right]    \cos \left( \omega t + q y \right)   \exp \left( - t/\tau \right)
\end{split}
\label{eq:fit}
\end{align}

\noindent The amplitude $A_0$, $v_\mathrm{dr}$, the diffusion constant $D_\mathrm{s}$, the dephasing time $\tau$, $\omega$ and the wavenumber $q$ are treated as fit parameters.
Detailed information on the fitting procedure is given in the Supplementary Information. 
To avoid deviations due to heating effects and other initial dynamics \cite{Henn2013a} not captured in this simple model, we fit the data from $t =300~\mathrm{ps}$. The decrease of the spatial precession period in time is a known effect of the finite size of the pump and probe laser spots~\cite{Salis2014}, and is accounted for by convolving Eq.\ (\ref{eq:fit}) with the Gaussian intensity profiles of the laser spots.
The experiment is perfectly described by this model, as evident from the good overlap of the symbols (experiment) with the solid lines (fits) in Fig.\ \ref{fig2}(a), and from the fitted gray lines that mark $S_z (y,t) = 0$ in the colorscale plots of Figs.\ \ref{fig2}(b-c). 

The fit parameters obtained for different values of $E_y$ are shown in Fig.\ \ref{fig3}. In Fig.\ \ref{fig3}(a), $v_\mathrm{dr}$ obtained from $S_z (y,t)$ is compared with values deduced from the measured current $I$ using $v_\mathrm{dr} = I /(e n_\mathrm{s} w)$, where $e$ is the elementary electron charge. The good agreement shows that the spin packet follows the stream of drifting electrons in the channel and that no parallel conductance obscures the interpretation of our data.
In Figs.\ \ref{fig3}(b-c), we summarize the values obtained for $q$ and $\omega$.
While $q$ shows no significant dependence on $v_\mathrm{dr}$, we find a linear dependence of $\omega$ on $v_\mathrm{dr}$ with a negative slope.

\begin{figure}
\includegraphics[width=\columnwidth]{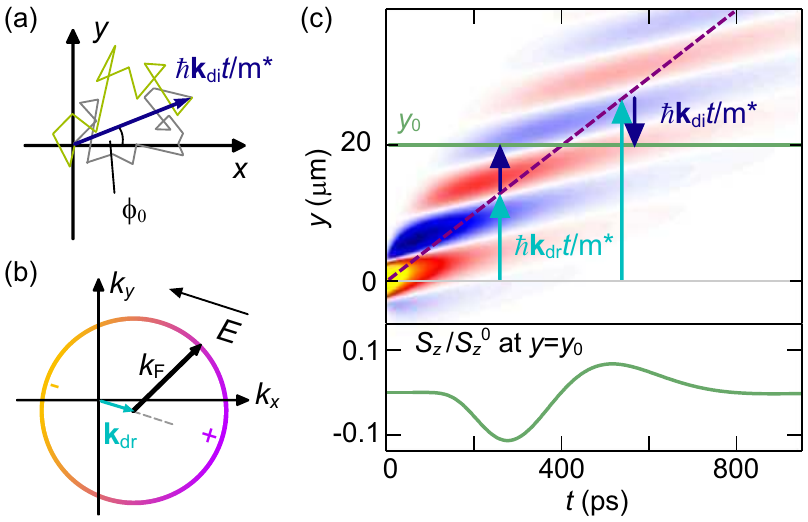}
\caption{Model of drift and diffusion. (a) Scattering events lead to diffusive trajectories of individual electrons. Shown are two trajectories of electrons that travel the same distance $\hbar \mathbf{k}_\mathrm{di} t / m^*$. (b) The Fermi circle is shifted by the drift vector $\mathbf{k}_\mathrm{dr}$.  (c) Exemplary map of $S_z (y,t)$ generated from Eq.\ (\ref{eq:fit}). Electrons with an average $\mathbf{k}_\mathrm{di} = 0$ drift along the violet dashed line. Electrons measured away from this line additionally experience a diffusive motion. Because of the unequal contributions of drift and diffusion to spin precession, the phase of quasi-stationary electron spins (for example, those on the solid green line) depends on how the travel is divided between drift and diffusion. This leads to a precession in time, as seen in the lower panel (shown for spins at $y = y_0$). 
\label{fig1} }
\end{figure}

Next, we show that the drift-induced $\omega$ is a consequence of cubic SOI.
Considering a degenerate 2DEG in a $(001)$-oriented quantum well with one occupied subband, the SOI field is given by \cite{Winkler, Walser2012}

\begin{equation}
\mathbf{\Omega}_\mathrm{SO} 
= \frac{2}{\hbar} 
\begin{pmatrix}
\left[ \alpha + \beta_1 + \beta_3 \frac{2 (k_x^2 - k_y^2)}{k_\mathrm{F}^2} \right] k_y\\
\left[- \alpha + \beta_1 + \beta_3 \frac{2 (k_y^2 - k_x^2)}{k_\mathrm{F}^2} \right] k_x
\end{pmatrix}
\, .
\label{eq:SOI}
\end{equation}

Here, $\alpha$ is the Rashba-coefficient, and $\beta_1$ and $\beta_3$ are the linear and cubic Dresselhaus coefficients, respectively. In the degenerate limit, the relevant electrons are those at the Fermi energy, $E_\mathrm{F} = \frac{\hbar^2 k_\mathrm{F}^2}{2 m^*}$, where $\hbar$ is the reduced Planck's constant, $m^*$ is the effective electron mass and $k_\mathrm{F}$ is the Fermi wave-vector.

Figure \ref{fig1}(a) sketches two different diffusive paths of electrons that travel the same distance  $\hbar \mathbf{k}_\mathrm{di} t / m^*$. On those paths, the electrons scatter many times and thereby sample different $k$-states. Because we consider electrons that travel along $\mathbf{k}_\mathrm{di}$, they occupy states with $k$-vectors along $\mathbf{k}_\mathrm{di}$ more often than along the opposite direction. Assuming isotropic scattering, this occupation is modeled by a weighting function 

\begin{equation}
f (\theta) = 1 + \frac{2 k_\mathrm{di}}{k_\mathrm{F}} \cos (\theta - \phi_0) \, ,
\end{equation}

\noindent such that the average momentum is $\hbar/(2\pi) \int^{2\pi}_{0} \mathbf{k} f(\theta)\,\mathrm{d}\theta = \hbar \mathbf{k}_\mathrm{di} $, with $\mathbf{k} = k (\cos \theta, \sin \theta)$ and $\mathbf{k}_\mathrm{di} = k_\mathrm{di} (\cos \phi_0, \sin \phi_0)$. 
The drift of the electron gas is accounted for by a shift of the Fermi circle by $\mathbf{k}_\mathrm{dr}$ [Fig.\ \ref{fig1}(b)]. Because of its dependence on $k$ [Eq.\ (\ref{eq:SOI})], the SOI field changes after each scattering event. Its average is given by $\langle \mathbf{\Omega}_\mathrm{SO} \rangle =  \int_0^{2\pi} d\theta \, \mathbf{\Omega}_\mathrm{SO} (\mathbf{k} + \mathbf{k}_\mathrm{dr}) \, f(\theta)$. Instead of deriving the spin mode of the system \cite{Yang2010}, we describe the spin dynamics by assuming that spins injected at $t=0$ and $x = y = 0$ precess about $\langle \mathbf{\Omega}_\mathrm{SO} \rangle$. 
For drift along the $y$-direction ($k_{\mathrm{dr},x} = 0$) and detection at $x = 0$ ($k_{\mathrm{di},x} = 0$), we obtain (see Supplementary Information for the general case)

\begin{equation}
\langle \mathbf{\Omega}_\mathrm{SO} \rangle = \frac{2 (\alpha + \beta_1)}{\hbar} (k_{\mathrm{di}, y} + k_{\mathrm{dr},y}) - \frac{2 \beta_3}{\hbar} \left( k_{\mathrm{di},y} + 2 k_{\mathrm{dr},y} \right) \, .
\label{eq:omgmean}
\end{equation}

This is a surprising result, because in the last term, which is proportional to $\beta_3$, drift ($k_{\mathrm{dr},y}$) leads to a spin precession angle twice as large as that induced by diffusion ($k_{\mathrm{di},y}$). 
As illustrated in Fig.\ \ref{fig1}(c), this leads to a precession in time for spins located at a constant position $y_0$. Without diffusion, the electrons follow $y = \hbar k_{\mathrm{dr},y} t / m^*$ (violet dashed line) and reach $y = y_0$ at a given time. Spins that reach $y_0$ earlier (later) will in addition diffuse along (against) $k_\mathrm{dr}$ and therefore acquire a different precession phase. To calculate the corresponding frequency $\omega$, we insert $y = \frac{\hbar}{m^*} (k_{\mathrm{dr},y} + k_{\mathrm{di},y}) t$ into Eq.\ (\ref{eq:omgmean}) and obtain $S_z(y, t) =  \cos\,  \langle \mathbf{\Omega}_\mathrm{SO} \rangle t = \cos \left( \omega t + q y \right)$, with 

\begin{align}
q  = &  \frac{2 m^*}{\hbar^2} (\alpha + \beta^*) \,\, \mathrm{and} \label{eq:q}\\ 
\omega = &  -\frac{2 m^*}{\hbar^2} v_\mathrm{dr} \beta_3 \, .
\label{eq:omg}
\end{align}

We have defined $\beta^* = \beta_1 - \beta_3$ and $v_\mathrm{dr} = \hbar k_\mathrm{dr} / m^*$. 
The wavenumber $q$ is not modified by drift to first order \footnote{Higher order corrections are discussed in the Supplementary Information.}. In contrast, the precession frequency $\omega$ depends linearly on $v_\mathrm{dr}$ and is proportional to the cubic Dresselhaus coefficient, $\beta_3$. This induces a temporal precession for quasi-stationary electrons [cf. lower panel in Fig.\ \ref{fig1}(c)]. The tilt of the green solid lines in Figs.\ \ref{fig2}(b-c) therefore directly visualizes the unequal contributions of drift and diffusion to the spin precession for nonlinear SOI. We note that spins that follow $y = v_\mathrm{dr}t$ precess with a frequency $\omega = \frac{2 m^*}{\hbar^2} v_\mathrm{dr} (\alpha + \beta_1 - 2\beta_3)$, recovering the result of Ref. [\onlinecite{Studer2010}], which is valid for measurements that do not spatially resolve the spin distribution. 

We find a remarkable agreement between Eqs.\ (\ref{eq:q}) and (\ref{eq:omg}) and the measured values for $q$ and $\omega$ [Figs.\ \ref{fig3}(b-c)]. From $q$, we obtain $\alpha + \beta^* = 6.2 \times 10^{-13}$~eVm, which is equal to previous results from a similar sample \cite{Salis2014}. The slope of $\omega$ vs.\ $v_\mathrm{dr}$ is directly proportional to $\beta_3$. We get $\beta_3 = 8.5 \times 10^{-14}$~eVm, which agrees perfectly with the measured sheet electron density of $n_\mathrm{s} = 5 \times 10^{15}$~m$^{-2}$ and a bulk Dresselhaus coefficient of $\gamma = - 11 \times 10^{-30}$~eVm$^3$~\cite{Walser2012PRB}, by considering that $\beta_3 = - \gamma \pi n_\mathrm{s} /2$.

\begin{figure}
\includegraphics[width=\columnwidth]{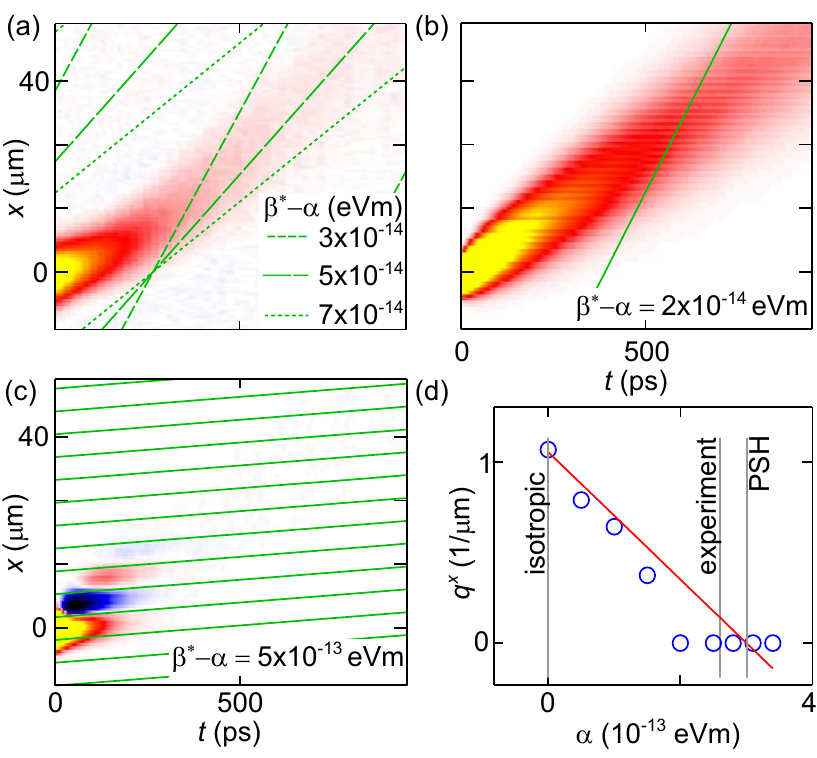}
\caption{Drift along $x$. (a) Measured $S_z (x, t)$ for $E_x = -3$~kV/m. No precession is visible, although for our sample we expect $3 \times 10^{-14} ~\mathrm{eVm} < \beta^* - \alpha < 7 \times 10^{-14} ~\mathrm{eVm}$, for which our model predicts a precession (green dashed lines mark $S_z = 0$). (b) Numerical Monte-Carlo simulation data of $S_z(x,t)$ for $\alpha = 2.9\times 10^{-13}$~eVm and $\beta^* = 3.1 \times 10^{-13}$~eVm. As in (a), no precession pattern is observed although it is predicted by the model (green solid line). (c) Numerical Monte-Carlo simulation data of $S_z(x,t)$ for $\alpha = 0.5\times 10^{-13}$~eVm and $\beta^* = 5.5 \times 10^{-13}$~eVm. Here, the SOI is almost isotropic and the model (green solid lines) describes the precession pattern well. (d) When $\alpha$ is gradually increased from zero, the $q_x$ observed in the simulation (blue circles) initially follows $q_x = 2 m^* (\beta^* - \alpha)/\hbar^2$ (red line). In a finite range around $\alpha = \beta^*$ (PSH), precession along $x$ is suppressed ($q_x = 0$). 
The total strength of SOI in all simulations [(b)-(d)] is kept constant at $\alpha + \beta^* = 6 \times 10^{-13}$~eVm.
\label{fig4} }
\end{figure}

Equations (\ref{eq:q}) and (\ref{eq:omg}) were derived assuming spin precession about an averaged $\langle \mathbf{\Omega}_\mathrm{SO} \rangle$. For drift along $y$, this is appropriate for the PSH situation ($\alpha = \beta^*$), where SOI is large for $\mathbf{k} || y$ and small for $\mathbf{k} || x$ [cf. Eq.\ (\ref{eq:SOI})]. The spin helix is described by a strong spatial spin precession along $y$ and no precession along $x$ \cite{Bernevig2006, Salis2014}. The investigated sample slightly deviates from the PSH symmetry, because $\beta^* - \alpha \neq 0$ as determined from measurements in an external magnetic field \cite{Walser2012, Salis2014, Altmann2014}: $3 \times 10^{-14}~\mathrm{eVm} < (\beta^* - \alpha) < 7 \times 10^{-14}~\mathrm{eVm}$. For drift along $x$, the model predicts a finite spatial spin precession with $q_x = 2 m^* (\beta^* - \alpha) / \hbar^2$. 
However, when we apply the electric field along the $x$ axis and measure $S_z (x,t)$, no precession is visible [Fig.\ \ref{fig4}(a)].
The absence of precession can be explained by the large anisotropy of the SOI. 
The small SOI field induced by drift along $x$ cannot destabilize the spin helix along $y$, which leads to the suppression of $q_x$. 
A similar effect has been predicted in a purely diffusive situation~\cite{Stanescu_2007, Poshakinskiy2015}. 
It is not accounted for in our simple model, where for drift along $x$, the fields for $\mathbf{k} || y$ average to zero and the fields induced by drift along $x$ appear dominant, even though electrons tracked at $y=0$ also occupy states with $\mathbf{k} || y$.

We compare the measured and modeled spin dynamics with a numerical Monte-Carlo simulation that takes the precession about all axes into account correctly. We set $\beta^* -\alpha =0.2 \times 10^{-13}$~eVm, $\beta_3 = 0.6 \times 10^{-13}$~eVm and $\alpha+\beta^* = 6 \times 10^{-13}$~eVm. Using Eq.\ (\ref{eq:SOI}), we calculate, in small time steps of 0.1\,ps, the traces of 500,000 electron spins that isotropically scatter on a Fermi circle (scattering time $\tau=0.7$~ps, $k_\mathrm{F}=1.6\times 10^8~\mathrm{m}^{-1}$) displaced along the $k_x$ direction by $k_\mathrm{dr} = 2.2 \times 10^7~\mathrm{m}^{-1}$. The result is shown in Fig.\ \ref{fig4}(b). As in the experiment, spin precession is absent.
In Fig.\ \ref{fig4}(c), the simulation data is shown for $\alpha=0.5$ and $\beta^*=5.5 \times 10^{-13}$~eVm. For this almost isotropic SOI~\cite{Altmann2015}, the model predicts both the temporal and the spatial precession period remarkably well (green lines). The transition from isotropic SOI to a PSH situation, for drift along $x$, is summarized in Fig.\ \ref{fig4}(d). It shows the wavenumber $q_x$ obtained from Monte-Carlo simulations as a function of $\alpha$. The value of $\beta^*$ was varied to keep $\alpha+\beta^*$ constant at $6 \times 10^{-13}$~eVm. The PSH situation is realized at $\alpha=3 \times 10^{-13}$~eVm, where the model correctly predicts $q_x=0$. Between there and $\alpha=2 \times 10^{-13}$~eVm, spin precession along $x$ is completely suppressed, in contrast to the linearly increasing $q_x$ of the simple model (red solid line). At smaller values of $\alpha$, towards the isotropic case, the simulated $q_x$ gradually approaches the model's prediction. 
In contrast, spin precession for drift along $y$ is correctly described by Eqs.\ (\ref{eq:q}) and (\ref{eq:omg}) for the entire range between the isotropic and the PSH case (see Supplementary Information). 
Note that in wire structures narrower than the SOI length, spin precession perpendicular to the wire is suppressed \cite{Malshukov2000, Kiselev2000, Altmann2015}, and we expect drift-induced spin precession to occur along the wire in any crystallographic direction for generic SOI.

In conclusion, we experimentally observed and theoretically explained that, for quasi-stationary electrons, current induces a temporal spin-precession frequency that is directly proportional to the drift velocity and the strength of cubic SOI. The origin of this effect is that drift motion in a cubic SOI system leads to a precession angle twice as large as that induced by diffusive motion. 
Further work is needed to analytically describe the spin precession for drift along the axis of weak SOI in an anisotropic situation. 
The occupation of a second subband or anisotropic scattering could modify the proportionality constant between $\omega$ and $\beta_3$. 
The temporal precession observed should hold universally for cubic SOI, e.g., also in hole gases in group IV \cite{Moriya2014} and III-V semiconductors \cite{Dyakonov1972, Winkler2002, Winkler, Minkov2005, Park2013}, or charge layers in oxides like perovskites \cite{Nakamura2012}. 
Moreover, the effect demonstrated must be considered when designing spintronic devices based on such systems. For read-out schemes with finite-sized contacts, it may lead to a temporal smearing of the spin packet and by that to signal reduction. This can be suppressed by designing a small diffusion constant. The effect itself presents a means to manipulate quasi-stationary spins via SOI and to directly quantify the strength of the cubic Dresselhaus SOI.

We acknowledge financial support from the NCCR QSIT of the Swiss National Science Foundation, F.G.G.H. acknowledges financial support from grants No. 2013/03450-7 and 2014/25981-7 of the S\~ao Paulo Research Foundation (FAPESP), G.J.F. acknowledges financial support from FAPEMIG and CNPq, and M.K. from the Japanese Ministry of Education, Culture, Sports, Science, and Technology (MEXT) in Grant-in-Aid for Scientific Research Nos. 15H02099 and 25220604. We thank R. Allenspach, A. Fuhrer, T. Henn, F. Valmorra, and R.~J. Warburton for helpful discussions, and U. Drechsler for technical assistance.

\FloatBarrier

\pagebreak
\onecolumngrid
\appendix

\section{Supplementary Information}

\subsection{Theory}

In Eq.\ (4) of the main text, we give the result of $\langle \mathbf{\Omega}_\mathrm{SO} \rangle  =  \int_0^{2\pi} d\theta \, \mathbf{\Omega}_\mathrm{SO} (\mathbf{k} + \mathbf{k}_\mathrm{dr}) \, f(\theta)$ for the special case that drift occurs along the $y$-direction ($k_{\mathrm{dr},x} = 0$) and detection at $x = 0$ ($k_{\mathrm{di},x} = 0$). 
Here, we provide the result for a general case:

\begin{equation}
\langle \mathbf{\Omega}_\mathrm{SO} \rangle  = \langle \mathbf{\Omega}_\mathrm{SO,1} \rangle  + \langle \mathbf{\Omega}_\mathrm{SO,3} \rangle \, ,
\end{equation}

with a term proportional to $\beta_1$:

\begin{equation}
\langle \mathbf{\Omega}_\mathrm{SO,1} \rangle  =  \frac{2 \beta_1}{\hbar} 
\begin{pmatrix} 
k_{\mathrm{di},y} + k_{\mathrm{dr},y} \\
k_{\mathrm{di},x} + k_{\mathrm{dr},x}
\end{pmatrix} \, ,
\end{equation}

and one proportional to $\beta_3$:

\begin{equation}
\langle \mathbf{\Omega}_\mathrm{SO,3} \rangle  =  \frac{2 \beta_3}{\hbar} 
\begin{pmatrix} 
- k_{\mathrm{di},y} -2 k_{\mathrm{dr},y} + \frac{2}{k_\mathrm{F}^2} \left[  k_{\mathrm{di},y} k_{\mathrm{dr},x}^2 + 2 k_{\mathrm{di},x} k_{\mathrm{dr},x}  k_{\mathrm{dr},y} +  k_{\mathrm{dr},x}^2 k_{\mathrm{dr},y} - 3 k_{\mathrm{di},y} k_{\mathrm{dr},y}^2 -  k_{\mathrm{dr},y}^3 \right] \\
- k_{\mathrm{di},x} -2 k_{\mathrm{dr},x} + \frac{2}{k_\mathrm{F}^2} \left[  k_{\mathrm{di},x} k_{\mathrm{dr},y}^2 + 2 k_{\mathrm{di},y} k_{\mathrm{dr},y}  k_{\mathrm{dr},x} +  k_{\mathrm{dr},y}^2 k_{\mathrm{dr},x} - 3 k_{\mathrm{di},x} k_{\mathrm{dr},x}^2 -  k_{\mathrm{dr},x}^3 \right]
\end{pmatrix}  \, .
\end{equation}

For simplicity, we assumed $\alpha = 0$ in the above expressions. 
We now move to the special case where the electric field is applied along the $y$ direction, such that $k_{\mathrm{dr},x} = 0$, and obtain

\begin{equation}
\langle \mathbf{\Omega}_\mathrm{SO,3} \rangle  = - \frac{2 \beta_3}{\hbar} 
\begin{pmatrix} 
k_{\mathrm{di},y} \left[1 + 6\left(\frac{k_{\mathrm{dr},y}}{k_\mathrm{F}}\right)^2 \right] + 2 k_{\mathrm{dr},y} \left[1 + \left(\frac{k_{\mathrm{dr},y}}{k_\mathrm{F}}\right)^2 \right] \\
k_{\mathrm{di},x} \left[ 1 - \left(\frac{2  k_{\mathrm{dr},y}}{k_\mathrm{F}} \right)^2 \right]
\end{pmatrix}  \, .
\end{equation}

To describe a measurement where spins are tracked along the drift direction $y$ and at $x = 0$, we additionally set $k_{\mathrm{di},x} = 0$ and obtain

\begin{equation}
\langle \mathbf{\Omega}_\mathrm{SO} \rangle  =   \frac{2 \beta_1}{\hbar} 
\begin{pmatrix} 
 k_{\mathrm{di},y} + k_{\mathrm{dr},y} \\
0
\end{pmatrix} 
- \frac{2 \beta_3}{\hbar} 
\begin{pmatrix} 
 k_{\mathrm{di},y} \left[1 + 6\left(\frac{k_{\mathrm{dr},y}}{k_\mathrm{F}}\right)^2 \right] + 2 k_{\mathrm{dr},y} \left[1 + \left(\frac{k_{\mathrm{dr},y}}{k_\mathrm{F}}\right)^2 \right]  \\
0
\end{pmatrix}  \, .
\end{equation}

Translating into a $(y, t)$ coordinate system via $y = \frac{\hbar}{m^*}t (k_{\mathrm{dr},y} + k_{\mathrm{di},y})$, we obtain

\begin{equation}
\langle \mathbf{\Omega}_\mathrm{SO} \rangle   t = q y + \omega t \, ,
\end{equation}

\noindent with

\begin{equation}
q = \frac{2 m^*}{\hbar^2} \left(  \beta_1 - \beta_3 \left[  1 + 6 \left(\frac{k_{\mathrm{dr},y}}{k_\mathrm{F}}\right)^2 \right] \right)
\end{equation}

\noindent and

\begin{equation}
\omega = -\frac{2 m^*}{\hbar^2} v_\mathrm{dr} \beta_3 \left[ 1 - 4 \left( \frac{k_{\mathrm{dr},y}}{k_\mathrm{F}} \right)^2  \right] \, .
\end{equation}

The higher-order terms are not observed in the experiment and therefore not discussed in the main text. However, they might be observable for larger drift vectors or smaller electron sheet densities.

\subsection{Simulations}

\begin{figure}
\includegraphics{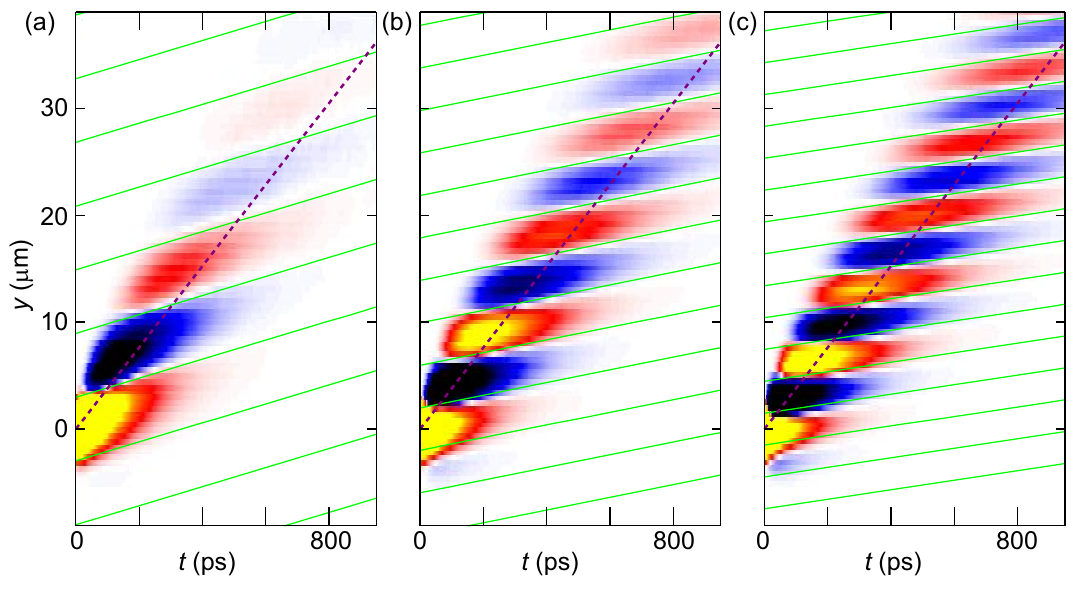}
\caption{Validity of the model. (a) to (c) Simulation data of $S_z(y,t)$ for $\alpha = 0$~eVm, $\alpha = 1.5 \times 10^{-13}$~eVm and $\alpha = 3 \times 10^{-13}$~eVm, respectively. In all cases, $\beta^* = 3 \times 10^{-13}$~eVm, $\beta_3 = 0.7 \times 10^{-13}$~eVm and $v_\mathrm{dr} = 38$~km/s (violet dashed line). We find good agreement between the simulation and the model (green solid lines) for the entire parameter range.
\label{SI_fig1} }
\end{figure}

In the main text, we discuss the validity of the model for cases away from the PSH symmetry, i.e., away from $\alpha = \beta^*$, by comparing the model with spin-precession maps obtained from numerical Monte-Carlo simulations. We state that, as long as drift occurs along $y$, we obtain good agreement between simulation and model. 
In Fig.\ \ref{SI_fig1}, we show the corresponding simulations for three different cases between $\alpha /\beta^* = 0$ (isotropic) and $\alpha /\beta^* = 1$ (PSH). The model of Eqs.\ (5) and (6) of the main text (green solid lines) correctly predicts the simulated spin dynamics for the entire parameter range for drift along $y$.

\subsection{Fitting}

Equation (1) in the main text contains six independent fit parameters. Suitable starting values for the fitting are obtained in the following way. For the amplitude $A_0$ we choose the value of $S_z (t=0, y=0)$. The drift velocity, $v_\mathrm{dr}$, is defined by the shift of the spin packet in time and its starting value is estimated manually. The spin diffusion constant, $D_\mathrm{s}$, is determined by the broadening of the Gaussian envelope function and we start with a typical value for samples from the same wafer. For the dephasing time, $\tau$, we use 1~ns as a starting value. The most important parameters for the presented study are $\omega$, the temporal precession frequency, and $q$, the spatial wavenumber. Both quantities are little affected by the other fit parameters. Starting values for both of them are obtained from a line-cut through the data at a fixed time (a fixed position) for $q$ (for $\omega$). Before calculating the mean-squared error between Eq. (1) and the measured $S_z (y, t)$, we perform a one-dimensional convolution of Eq. (1) with the Gaussian intensity profiles of the pump and probe laser spots along $y$. This step is very important, because its neglect distorts particularly the value of $\omega$.

All fit parameters are then constrained to a reasonable range. To determine each parameter's fit value and confidence interval, we vary that parameter in small steps through its full range. At each step, all other parameters are optimized to minimize the mean-squared error between the data and Eq. (1) by a Nelder-Mead simplex search algorithm. The value of the parameter with the smallest error defines the fit value. For all fit parameters, we find a single minimum. The confidence interval, as shown in Fig. 2 in the main text, is then defined by an increase of the mean-squared error by 5~\% from its minimal value. The mean-squared error is taken over approximately 3000 data points (typically 35 steps of $t$, 85 steps of $y$ or $x$).


\begin{thebibliography}{37}%
\makeatletter
\providecommand \@ifxundefined [1]{%
 \@ifx{#1\undefined}
}%
\providecommand \@ifnum [1]{%
 \ifnum #1\expandafter \@firstoftwo
 \else \expandafter \@secondoftwo
 \fi
}%
\providecommand \@ifx [1]{%
 \ifx #1\expandafter \@firstoftwo
 \else \expandafter \@secondoftwo
 \fi
}%
\providecommand \natexlab [1]{#1}%
\providecommand \enquote  [1]{``#1''}%
\providecommand \bibnamefont  [1]{#1}%
\providecommand \bibfnamefont [1]{#1}%
\providecommand \citenamefont [1]{#1}%
\providecommand \href@noop [0]{\@secondoftwo}%
\providecommand \href [0]{\begingroup \@sanitize@url \@href}%
\providecommand \@href[1]{\@@startlink{#1}\@@href}%
\providecommand \@@href[1]{\endgroup#1\@@endlink}%
\providecommand \@sanitize@url [0]{\catcode `\\12\catcode `\$12\catcode
  `\&12\catcode `\#12\catcode `\^12\catcode `\_12\catcode `\%12\relax}%
\providecommand \@@startlink[1]{}%
\providecommand \@@endlink[0]{}%
\providecommand \url  [0]{\begingroup\@sanitize@url \@url }%
\providecommand \@url [1]{\endgroup\@href {#1}{\urlprefix }}%
\providecommand \urlprefix  [0]{URL }%
\providecommand \Eprint [0]{\href }%
\providecommand \doibase [0]{http://dx.doi.org/}%
\providecommand \selectlanguage [0]{\@gobble}%
\providecommand \bibinfo  [0]{\@secondoftwo}%
\providecommand \bibfield  [0]{\@secondoftwo}%
\providecommand \translation [1]{[#1]}%
\providecommand \BibitemOpen [0]{}%
\providecommand \bibitemStop [0]{}%
\providecommand \bibitemNoStop [0]{.\EOS\space}%
\providecommand \EOS [0]{\spacefactor3000\relax}%
\providecommand \BibitemShut  [1]{\csname bibitem#1\endcsname}%
\let\auto@bib@innerbib\@empty
\bibitem [{\citenamefont {\ifmmode \check{Z}\else
  \v{Z}\fi{}uti\ifmmode~\acute{c}\else \'{c}\fi{}}\ \emph
  {et~al.}(2004)\citenamefont {\ifmmode \check{Z}\else
  \v{Z}\fi{}uti\ifmmode~\acute{c}\else \'{c}\fi{}}, \citenamefont {Fabian},\
  and\ \citenamefont {Das~Sarma}}]{ZuticFabianDasSarma}%
  \BibitemOpen
  \bibfield  {author} {\bibinfo {author} {\bibfnamefont {I.}~\bibnamefont
  {\ifmmode \check{Z}\else \v{Z}\fi{}uti\ifmmode~\acute{c}\else \'{c}\fi{}}},
  \bibinfo {author} {\bibfnamefont {J.}~\bibnamefont {Fabian}}, \ and\ \bibinfo
  {author} {\bibfnamefont {S.}~\bibnamefont {Das~Sarma}},\ }\href {\doibase
  10.1103/RevModPhys.76.323} {\bibfield  {journal} {\bibinfo  {journal} {Rev.
  Mod. Phys.}\ }\textbf {\bibinfo {volume} {76}},\ \bibinfo {pages} {323}
  (\bibinfo {year} {2004})}\BibitemShut {NoStop}%
\bibitem [{\citenamefont {Awschalom}\ and\ \citenamefont
  {Flatt\'{e}}(2007)}]{Awschalom2007}%
  \BibitemOpen
  \bibfield  {author} {\bibinfo {author} {\bibfnamefont {D.~D.}\ \bibnamefont
  {Awschalom}}\ and\ \bibinfo {author} {\bibfnamefont {M.~E.}\ \bibnamefont
  {Flatt\'{e}}},\ }\href@noop {} {\bibfield  {journal} {\bibinfo  {journal}
  {Nature Physics}\ }\textbf {\bibinfo {volume} {\textbf{3}}},\ \bibinfo
  {pages} {153} (\bibinfo {year} {2007})}\BibitemShut {NoStop}%
\bibitem [{\citenamefont {Behin-Aein}\ \emph {et~al.}(2010)\citenamefont
  {Behin-Aein}, \citenamefont {Datta}, \citenamefont {Salahuddin},\ and\
  \citenamefont {Datta}}]{Datta2010}%
  \BibitemOpen
  \bibfield  {author} {\bibinfo {author} {\bibfnamefont {B.}~\bibnamefont
  {Behin-Aein}}, \bibinfo {author} {\bibfnamefont {D.}~\bibnamefont {Datta}},
  \bibinfo {author} {\bibfnamefont {S.}~\bibnamefont {Salahuddin}}, \ and\
  \bibinfo {author} {\bibfnamefont {S.}~\bibnamefont {Datta}},\ }\href@noop {}
  {\bibfield  {journal} {\bibinfo  {journal} {Nature Nanotechnology}\ }\textbf
  {\bibinfo {volume} {\textbf{5}}},\ \bibinfo {pages} {266} (\bibinfo {year}
  {2010})}\BibitemShut {NoStop}%
\bibitem [{\citenamefont {{Winkler}}(2003)}]{Winkler}%
  \BibitemOpen
  \bibfield  {author} {\bibinfo {author} {\bibfnamefont {R.}~\bibnamefont
  {{Winkler}}},\ }\href@noop {} {\emph {\bibinfo {title} {Spin-{O}rbit
  {C}oupling {E}ffects in {T}wo-{D}imensional {E}lectron and {H}ole
  {S}ystems}}}\ (\bibinfo  {publisher} {Springer},\ \bibinfo {year}
  {2003})\BibitemShut {NoStop}%
\bibitem [{\citenamefont {{Datta}}\ and\ \citenamefont
  {{Das}}(1990)}]{Datta1990}%
  \BibitemOpen
  \bibfield  {author} {\bibinfo {author} {\bibfnamefont {S.}~\bibnamefont
  {{Datta}}}\ and\ \bibinfo {author} {\bibfnamefont {B.}~\bibnamefont
  {{Das}}},\ }\href@noop {} {\bibfield  {journal} {\bibinfo  {journal} {Appl.
  Phys. Lett.}\ }\textbf {\bibinfo {volume} {\textbf{56}}},\ \bibinfo {pages}
  {665} (\bibinfo {year} {1990})}\BibitemShut {NoStop}%
\bibitem [{\citenamefont {{Chuang}}\ \emph {et~al.}(2014)\citenamefont
  {{Chuang}}, \citenamefont {{Ho}}, \citenamefont {{Smith}}, \citenamefont
  {{Sfigakis}}, \citenamefont {{Pepper}}, \citenamefont {{Chen}}, \citenamefont
  {{Fan}}, \citenamefont {{Griffiths}}, \citenamefont {{Farrer}}, \citenamefont
  {{Beere}}, \citenamefont {{Jones}}, \citenamefont {{Ritchie}},\ and\
  \citenamefont {{Chen}}}]{Chuang2014}%
  \BibitemOpen
  \bibfield  {author} {\bibinfo {author} {\bibfnamefont {P.}~\bibnamefont
  {{Chuang}}}, \bibinfo {author} {\bibfnamefont {S.-C.}\ \bibnamefont {{Ho}}},
  \bibinfo {author} {\bibfnamefont {L.~W.}\ \bibnamefont {{Smith}}}, \bibinfo
  {author} {\bibfnamefont {F.}~\bibnamefont {{Sfigakis}}}, \bibinfo {author}
  {\bibfnamefont {M.}~\bibnamefont {{Pepper}}}, \bibinfo {author}
  {\bibfnamefont {C.-H.}\ \bibnamefont {{Chen}}}, \bibinfo {author}
  {\bibfnamefont {J.-C.}\ \bibnamefont {{Fan}}}, \bibinfo {author}
  {\bibfnamefont {J.~P.}\ \bibnamefont {{Griffiths}}}, \bibinfo {author}
  {\bibfnamefont {I.}~\bibnamefont {{Farrer}}}, \bibinfo {author}
  {\bibfnamefont {H.~E.}\ \bibnamefont {{Beere}}}, \bibinfo {author}
  {\bibfnamefont {G.~A.~C.}\ \bibnamefont {{Jones}}}, \bibinfo {author}
  {\bibfnamefont {D.~A.}\ \bibnamefont {{Ritchie}}}, \ and\ \bibinfo {author}
  {\bibfnamefont {T.-M.}\ \bibnamefont {{Chen}}},\ }\href@noop {} {\bibfield
  {journal} {\bibinfo  {journal} {Nature Nanotechnology}\ }\textbf {\bibinfo
  {volume} {\textbf{10}}},\ \bibinfo {pages} {35} (\bibinfo {year}
  {2014})}\BibitemShut {NoStop}%
\bibitem [{\citenamefont {Cameron}\ \emph {et~al.}(1996)\citenamefont
  {Cameron}, \citenamefont {Riblet},\ and\ \citenamefont
  {Miller}}]{Cameron1996}%
  \BibitemOpen
  \bibfield  {author} {\bibinfo {author} {\bibfnamefont {A.~R.}\ \bibnamefont
  {Cameron}}, \bibinfo {author} {\bibfnamefont {P.}~\bibnamefont {Riblet}}, \
  and\ \bibinfo {author} {\bibfnamefont {A.}~\bibnamefont {Miller}},\ }\href
  {\doibase 10.1103/PhysRevLett.76.4793} {\bibfield  {journal} {\bibinfo
  {journal} {Phys. Rev. Lett.}\ }\textbf {\bibinfo {volume} {76}},\ \bibinfo
  {pages} {4793} (\bibinfo {year} {1996})}\BibitemShut {NoStop}%
\bibitem [{\citenamefont {Kikkawa}\ and\ \citenamefont
  {Awschalom}(1999)}]{Kikkawa1999}%
  \BibitemOpen
  \bibfield  {author} {\bibinfo {author} {\bibfnamefont {J.~M.}\ \bibnamefont
  {Kikkawa}}\ and\ \bibinfo {author} {\bibfnamefont {D.~D.}\ \bibnamefont
  {Awschalom}},\ }\href
  {http://www.nature.com/nature/journal/v397/n6715/abs/397139a0.html}
  {\bibfield  {journal} {\bibinfo  {journal} {Nature}\ }\textbf {\bibinfo
  {volume} {397}},\ \bibinfo {pages} {139} (\bibinfo {year}
  {1999})}\BibitemShut {NoStop}%
\bibitem [{\citenamefont {Crooker}\ and\ \citenamefont
  {Smith}(2005)}]{Crooker2005}%
  \BibitemOpen
  \bibfield  {author} {\bibinfo {author} {\bibfnamefont {S.~A.}\ \bibnamefont
  {Crooker}}\ and\ \bibinfo {author} {\bibfnamefont {D.~L.}\ \bibnamefont
  {Smith}},\ }\href {\doibase 10.1103/PhysRevLett.94.236601} {\bibfield
  {journal} {\bibinfo  {journal} {Phys. Rev. Lett.}\ }\textbf {\bibinfo
  {volume} {94}},\ \bibinfo {pages} {236601} (\bibinfo {year}
  {2005})}\BibitemShut {NoStop}%
\bibitem [{\citenamefont {Wunderlich}\ \emph {et~al.}(2010)\citenamefont
  {Wunderlich}, \citenamefont {Park}, \citenamefont {Irvine}, \citenamefont
  {Zarbo}, \citenamefont {RozkotovÃ¡}, \citenamefont {Nemec}, \citenamefont
  {NovÃ¡k}, \citenamefont {Sinova},\ and\ \citenamefont
  {Jungwirth}}]{Wunderlich2010}%
  \BibitemOpen
  \bibfield  {author} {\bibinfo {author} {\bibfnamefont {J.}~\bibnamefont
  {Wunderlich}}, \bibinfo {author} {\bibfnamefont {B.-G.}\ \bibnamefont
  {Park}}, \bibinfo {author} {\bibfnamefont {A.~C.}\ \bibnamefont {Irvine}},
  \bibinfo {author} {\bibfnamefont {L.~P.}\ \bibnamefont {Zarbo}}, \bibinfo
  {author} {\bibfnamefont {E.}~\bibnamefont {RozkotovÃ¡}}, \bibinfo {author}
  {\bibfnamefont {P.}~\bibnamefont {Nemec}}, \bibinfo {author} {\bibfnamefont
  {V.}~\bibnamefont {NovÃ¡k}}, \bibinfo {author} {\bibfnamefont
  {J.}~\bibnamefont {Sinova}}, \ and\ \bibinfo {author} {\bibfnamefont
  {T.}~\bibnamefont {Jungwirth}},\ }\href {\doibase 10.1126/science.1195816}
  {\bibfield  {journal} {\bibinfo  {journal} {Science}\ }\textbf {\bibinfo
  {volume} {330}},\ \bibinfo {pages} {1801} (\bibinfo {year}
  {2010})}\BibitemShut {NoStop}%
\bibitem [{\citenamefont {V\"olkl}\ \emph {et~al.}(2011)\citenamefont
  {V\"olkl}, \citenamefont {Griesbeck}, \citenamefont {Tarasenko},
  \citenamefont {Schuh}, \citenamefont {Wegscheider}, \citenamefont
  {Sch\"uller},\ and\ \citenamefont {Korn}}]{Voelkl2011}%
  \BibitemOpen
  \bibfield  {author} {\bibinfo {author} {\bibfnamefont {R.}~\bibnamefont
  {V\"olkl}}, \bibinfo {author} {\bibfnamefont {M.}~\bibnamefont {Griesbeck}},
  \bibinfo {author} {\bibfnamefont {S.~A.}\ \bibnamefont {Tarasenko}}, \bibinfo
  {author} {\bibfnamefont {D.}~\bibnamefont {Schuh}}, \bibinfo {author}
  {\bibfnamefont {W.}~\bibnamefont {Wegscheider}}, \bibinfo {author}
  {\bibfnamefont {C.}~\bibnamefont {Sch\"uller}}, \ and\ \bibinfo {author}
  {\bibfnamefont {T.}~\bibnamefont {Korn}},\ }\href {\doibase
  10.1103/PhysRevB.83.241306} {\bibfield  {journal} {\bibinfo  {journal} {Phys.
  Rev. B}\ }\textbf {\bibinfo {volume} {83}},\ \bibinfo {pages} {241306}
  (\bibinfo {year} {2011})}\BibitemShut {NoStop}%
\bibitem [{\citenamefont {Dresselhaus}\ \emph {et~al.}(1992)\citenamefont
  {Dresselhaus}, \citenamefont {Papavassiliou}, \citenamefont {Wheeler},\ and\
  \citenamefont {Sacks}}]{Dresselhaus1992}%
  \BibitemOpen
  \bibfield  {author} {\bibinfo {author} {\bibfnamefont {P.~D.}\ \bibnamefont
  {Dresselhaus}}, \bibinfo {author} {\bibfnamefont {C.~M.~A.}\ \bibnamefont
  {Papavassiliou}}, \bibinfo {author} {\bibfnamefont {R.~G.}\ \bibnamefont
  {Wheeler}}, \ and\ \bibinfo {author} {\bibfnamefont {R.~N.}\ \bibnamefont
  {Sacks}},\ }\href {\doibase 10.1103/PhysRevLett.68.106} {\bibfield  {journal}
  {\bibinfo  {journal} {Phys. Rev. Lett.}\ }\textbf {\bibinfo {volume} {68}},\
  \bibinfo {pages} {106} (\bibinfo {year} {1992})}\BibitemShut {NoStop}%
\bibitem [{\citenamefont {Lou}\ \emph {et~al.}(2007)\citenamefont {Lou},
  \citenamefont {Adelmann}, \citenamefont {Crooker}, \citenamefont {Garlid},
  \citenamefont {Zhang}, \citenamefont {Reddy}, \citenamefont {Flexner},
  \citenamefont {Palmstr{\o}m},\ and\ \citenamefont {Crowell}}]{Lou2007}%
  \BibitemOpen
  \bibfield  {author} {\bibinfo {author} {\bibfnamefont {X.}~\bibnamefont
  {Lou}}, \bibinfo {author} {\bibfnamefont {C.}~\bibnamefont {Adelmann}},
  \bibinfo {author} {\bibfnamefont {S.~A.}\ \bibnamefont {Crooker}}, \bibinfo
  {author} {\bibfnamefont {E.~S.}\ \bibnamefont {Garlid}}, \bibinfo {author}
  {\bibfnamefont {J.}~\bibnamefont {Zhang}}, \bibinfo {author} {\bibfnamefont
  {K.~S.~M.}\ \bibnamefont {Reddy}}, \bibinfo {author} {\bibfnamefont {S.~D.}\
  \bibnamefont {Flexner}}, \bibinfo {author} {\bibfnamefont {C.~J.}\
  \bibnamefont {Palmstr{\o}m}}, \ and\ \bibinfo {author} {\bibfnamefont
  {P.~A.}\ \bibnamefont {Crowell}},\ }\href@noop {} {\bibfield  {journal}
  {\bibinfo  {journal} {Nature Physics}\ }\textbf {\bibinfo {volume} {3}},\
  \bibinfo {pages} {197} (\bibinfo {year} {2007})}\BibitemShut {NoStop}%
\bibitem [{\citenamefont {Stanescu}\ and\ \citenamefont
  {Galitski}(2007)}]{Stanescu_2007}%
  \BibitemOpen
  \bibfield  {author} {\bibinfo {author} {\bibfnamefont {T.~D.}\ \bibnamefont
  {Stanescu}}\ and\ \bibinfo {author} {\bibfnamefont {V.}~\bibnamefont
  {Galitski}},\ }\href {\doibase 10.1103/PhysRevB.75.125307} {\bibfield
  {journal} {\bibinfo  {journal} {Phys. Rev. B}\ }\textbf {\bibinfo {volume}
  {75}},\ \bibinfo {pages} {125307} (\bibinfo {year} {2007})}\BibitemShut
  {NoStop}%
\bibitem [{\citenamefont {Hru\ifmmode~\check{s}\else \v{s}\fi{}ka}\ \emph
  {et~al.}(2006)\citenamefont {Hru\ifmmode~\check{s}\else \v{s}\fi{}ka},
  \citenamefont {Kos}, \citenamefont {Crooker}, \citenamefont {Saxena},\ and\
  \citenamefont {Smith}}]{Hruska2006}%
  \BibitemOpen
  \bibfield  {author} {\bibinfo {author} {\bibfnamefont {M.}~\bibnamefont
  {Hru\ifmmode~\check{s}\else \v{s}\fi{}ka}}, \bibinfo {author} {\bibfnamefont
  {v.}~\bibnamefont {Kos}}, \bibinfo {author} {\bibfnamefont {S.~A.}\
  \bibnamefont {Crooker}}, \bibinfo {author} {\bibfnamefont {A.}~\bibnamefont
  {Saxena}}, \ and\ \bibinfo {author} {\bibfnamefont {D.~L.}\ \bibnamefont
  {Smith}},\ }\href {\doibase 10.1103/PhysRevB.73.075306} {\bibfield  {journal}
  {\bibinfo  {journal} {Phys. Rev. B}\ }\textbf {\bibinfo {volume} {73}},\
  \bibinfo {pages} {075306} (\bibinfo {year} {2006})}\BibitemShut {NoStop}%
\bibitem [{\citenamefont {Cheng}\ and\ \citenamefont {Wu}(2007)}]{Cheng2007}%
  \BibitemOpen
  \bibfield  {author} {\bibinfo {author} {\bibfnamefont {J.~L.}\ \bibnamefont
  {Cheng}}\ and\ \bibinfo {author} {\bibfnamefont {M.~W.}\ \bibnamefont {Wu}},\
  }\href {\doibase http://dx.doi.org/10.1063/1.2717526} {\bibfield  {journal}
  {\bibinfo  {journal} {Journal of Applied Physics}\ }\textbf {\bibinfo
  {volume} {101}},\ \bibinfo {eid} {073702} (\bibinfo {year} {2007}),\
  http://dx.doi.org/10.1063/1.2717526}\BibitemShut {NoStop}%
\bibitem [{\citenamefont {Yang}\ \emph {et~al.}(2010)\citenamefont {Yang},
  \citenamefont {Orenstein},\ and\ \citenamefont {Lee}}]{Yang2010}%
  \BibitemOpen
  \bibfield  {author} {\bibinfo {author} {\bibfnamefont {L.}~\bibnamefont
  {Yang}}, \bibinfo {author} {\bibfnamefont {J.}~\bibnamefont {Orenstein}}, \
  and\ \bibinfo {author} {\bibfnamefont {D.-H.}\ \bibnamefont {Lee}},\ }\href
  {\doibase 10.1103/PhysRevB.82.155324} {\bibfield  {journal} {\bibinfo
  {journal} {Phys. Rev. B}\ }\textbf {\bibinfo {volume} {82}},\ \bibinfo
  {pages} {155324} (\bibinfo {year} {2010})}\BibitemShut {NoStop}%
\bibitem [{\citenamefont {Yang}\ \emph {et~al.}(2012)\citenamefont {Yang},
  \citenamefont {Koralek}, \citenamefont {Orenstein}, \citenamefont {Tibbetts},
  \citenamefont {Reno},\ and\ \citenamefont {Lilly}}]{Yang2012}%
  \BibitemOpen
  \bibfield  {author} {\bibinfo {author} {\bibfnamefont {L.}~\bibnamefont
  {Yang}}, \bibinfo {author} {\bibfnamefont {J.~D.}\ \bibnamefont {Koralek}},
  \bibinfo {author} {\bibfnamefont {J.}~\bibnamefont {Orenstein}}, \bibinfo
  {author} {\bibfnamefont {D.~R.}\ \bibnamefont {Tibbetts}}, \bibinfo {author}
  {\bibfnamefont {J.~L.}\ \bibnamefont {Reno}}, \ and\ \bibinfo {author}
  {\bibfnamefont {M.~P.}\ \bibnamefont {Lilly}},\ }\href {\doibase
  10.1103/PhysRevLett.109.246603} {\bibfield  {journal} {\bibinfo  {journal}
  {Phys. Rev. Lett.}\ }\textbf {\bibinfo {volume} {109}},\ \bibinfo {pages}
  {246603} (\bibinfo {year} {2012})}\BibitemShut {NoStop}%
\bibitem [{\citenamefont {Schliemann}\ \emph {et~al.}(2003)\citenamefont
  {Schliemann}, \citenamefont {Egues},\ and\ \citenamefont
  {Loss}}]{Schliemann2003}%
  \BibitemOpen
  \bibfield  {author} {\bibinfo {author} {\bibfnamefont {J.}~\bibnamefont
  {Schliemann}}, \bibinfo {author} {\bibfnamefont {J.~C.}\ \bibnamefont
  {Egues}}, \ and\ \bibinfo {author} {\bibfnamefont {D.}~\bibnamefont {Loss}},\
  }\href {\doibase 10.1103/PhysRevLett.90.146801} {\bibfield  {journal}
  {\bibinfo  {journal} {Phys. Rev. Lett.}\ }\textbf {\bibinfo {volume} {90}},\
  \bibinfo {pages} {146801} (\bibinfo {year} {2003})}\BibitemShut {NoStop}%
\bibitem [{\citenamefont {Bernevig}\ \emph {et~al.}(2006)\citenamefont
  {Bernevig}, \citenamefont {Orenstein},\ and\ \citenamefont
  {Zhang}}]{Bernevig2006}%
  \BibitemOpen
  \bibfield  {author} {\bibinfo {author} {\bibfnamefont {B.~A.}\ \bibnamefont
  {Bernevig}}, \bibinfo {author} {\bibfnamefont {J.}~\bibnamefont {Orenstein}},
  \ and\ \bibinfo {author} {\bibfnamefont {S.-C.}\ \bibnamefont {Zhang}},\
  }\href {\doibase 10.1103/PhysRevLett.97.236601} {\bibfield  {journal}
  {\bibinfo  {journal} {Phys. Rev. Lett.}\ }\textbf {\bibinfo {volume} {97}},\
  \bibinfo {pages} {236601} (\bibinfo {year} {2006})}\BibitemShut {NoStop}%
\bibitem [{\citenamefont {{Walser}}\ \emph
  {et~al.}(2012{\natexlab{a}})\citenamefont {{Walser}}, \citenamefont
  {{Reichl}}, \citenamefont {{Wegscheider}},\ and\ \citenamefont
  {{Salis}}}]{Walser2012}%
  \BibitemOpen
  \bibfield  {author} {\bibinfo {author} {\bibfnamefont {M.~P.}\ \bibnamefont
  {{Walser}}}, \bibinfo {author} {\bibfnamefont {C.}~\bibnamefont {{Reichl}}},
  \bibinfo {author} {\bibfnamefont {W.}~\bibnamefont {{Wegscheider}}}, \ and\
  \bibinfo {author} {\bibfnamefont {G.}~\bibnamefont {{Salis}}},\ }\href
  {http://www.nature.com/nphys/journal/v8/n10/abs/nphys2383.html} {\bibfield
  {journal} {\bibinfo  {journal} {Nat. Phys.}\ }\textbf {\bibinfo {volume}
  {\textbf{8}}},\ \bibinfo {pages} {757} (\bibinfo {year}
  {2012}{\natexlab{a}})}\BibitemShut {NoStop}%
\bibitem [{\citenamefont {Altmann}\ \emph {et~al.}(2014)\citenamefont
  {Altmann}, \citenamefont {Walser}, \citenamefont {Reichl}, \citenamefont
  {Wegscheider},\ and\ \citenamefont {Salis}}]{Altmann2014}%
  \BibitemOpen
  \bibfield  {author} {\bibinfo {author} {\bibfnamefont {P.}~\bibnamefont
  {Altmann}}, \bibinfo {author} {\bibfnamefont {M.~P.}\ \bibnamefont {Walser}},
  \bibinfo {author} {\bibfnamefont {C.}~\bibnamefont {Reichl}}, \bibinfo
  {author} {\bibfnamefont {W.}~\bibnamefont {Wegscheider}}, \ and\ \bibinfo
  {author} {\bibfnamefont {G.}~\bibnamefont {Salis}},\ }\href {\doibase
  10.1103/PhysRevB.90.201306} {\bibfield  {journal} {\bibinfo  {journal} {Phys.
  Rev. B}\ }\textbf {\bibinfo {volume} {90}},\ \bibinfo {pages} {201306}
  (\bibinfo {year} {2014})}\BibitemShut {NoStop}%
\bibitem [{\citenamefont {Altmann}\ \emph {et~al.}(2015)\citenamefont
  {Altmann}, \citenamefont {Kohda}, \citenamefont {Reichl}, \citenamefont
  {Wegscheider},\ and\ \citenamefont {Salis}}]{Altmann2015}%
  \BibitemOpen
  \bibfield  {author} {\bibinfo {author} {\bibfnamefont {P.}~\bibnamefont
  {Altmann}}, \bibinfo {author} {\bibfnamefont {M.}~\bibnamefont {Kohda}},
  \bibinfo {author} {\bibfnamefont {C.}~\bibnamefont {Reichl}}, \bibinfo
  {author} {\bibfnamefont {W.}~\bibnamefont {Wegscheider}}, \ and\ \bibinfo
  {author} {\bibfnamefont {G.}~\bibnamefont {Salis}},\ }\href {\doibase
  10.1103/PhysRevB.92.235304} {\bibfield  {journal} {\bibinfo  {journal} {Phys.
  Rev. B}\ }\textbf {\bibinfo {volume} {92}},\ \bibinfo {pages} {235304}
  (\bibinfo {year} {2015})}\BibitemShut {NoStop}%
\bibitem [{\citenamefont {Henn}\ \emph {et~al.}(2013)\citenamefont {Henn},
  \citenamefont {Heckel}, \citenamefont {Beck}, \citenamefont {Kiessling},
  \citenamefont {Ossau}, \citenamefont {Molenkamp}, \citenamefont {Reuter},\
  and\ \citenamefont {Wieck}}]{Henn2013a}%
  \BibitemOpen
  \bibfield  {author} {\bibinfo {author} {\bibfnamefont {T.}~\bibnamefont
  {Henn}}, \bibinfo {author} {\bibfnamefont {A.}~\bibnamefont {Heckel}},
  \bibinfo {author} {\bibfnamefont {M.}~\bibnamefont {Beck}}, \bibinfo {author}
  {\bibfnamefont {T.}~\bibnamefont {Kiessling}}, \bibinfo {author}
  {\bibfnamefont {W.}~\bibnamefont {Ossau}}, \bibinfo {author} {\bibfnamefont
  {L.~W.}\ \bibnamefont {Molenkamp}}, \bibinfo {author} {\bibfnamefont
  {D.}~\bibnamefont {Reuter}}, \ and\ \bibinfo {author} {\bibfnamefont {A.~D.}\
  \bibnamefont {Wieck}},\ }\href {\doibase 10.1103/PhysRevB.88.085303}
  {\bibfield  {journal} {\bibinfo  {journal} {Phys. Rev. B}\ }\textbf {\bibinfo
  {volume} {88}},\ \bibinfo {pages} {085303} (\bibinfo {year}
  {2013})}\BibitemShut {NoStop}%
\bibitem [{\citenamefont {Salis}\ \emph {et~al.}(2014)\citenamefont {Salis},
  \citenamefont {Walser}, \citenamefont {Altmann}, \citenamefont {Reichl},\
  and\ \citenamefont {Wegscheider}}]{Salis2014}%
  \BibitemOpen
  \bibfield  {author} {\bibinfo {author} {\bibfnamefont {G.}~\bibnamefont
  {Salis}}, \bibinfo {author} {\bibfnamefont {M.~P.}\ \bibnamefont {Walser}},
  \bibinfo {author} {\bibfnamefont {P.}~\bibnamefont {Altmann}}, \bibinfo
  {author} {\bibfnamefont {C.}~\bibnamefont {Reichl}}, \ and\ \bibinfo {author}
  {\bibfnamefont {W.}~\bibnamefont {Wegscheider}},\ }\href {\doibase
  10.1103/PhysRevB.89.045304} {\bibfield  {journal} {\bibinfo  {journal} {Phys.
  Rev. B}\ }\textbf {\bibinfo {volume} {89}},\ \bibinfo {pages} {045304}
  (\bibinfo {year} {2014})}\BibitemShut {NoStop}%
\bibitem [{Note1()}]{Note1}%
  \BibitemOpen
  \bibinfo {note} {Higher order corrections are discussed in the Supplementary
  Information.}\BibitemShut {Stop}%
\bibitem [{\citenamefont {Studer}\ \emph {et~al.}(2010)\citenamefont {Studer},
  \citenamefont {Walser}, \citenamefont {Baer}, \citenamefont {Rusterholz},
  \citenamefont {Sch\"on}, \citenamefont {Schuh}, \citenamefont {Wegscheider},
  \citenamefont {Ensslin},\ and\ \citenamefont {Salis}}]{Studer2010}%
  \BibitemOpen
  \bibfield  {author} {\bibinfo {author} {\bibfnamefont {M.}~\bibnamefont
  {Studer}}, \bibinfo {author} {\bibfnamefont {M.~P.}\ \bibnamefont {Walser}},
  \bibinfo {author} {\bibfnamefont {S.}~\bibnamefont {Baer}}, \bibinfo {author}
  {\bibfnamefont {H.}~\bibnamefont {Rusterholz}}, \bibinfo {author}
  {\bibfnamefont {S.}~\bibnamefont {Sch\"on}}, \bibinfo {author} {\bibfnamefont
  {D.}~\bibnamefont {Schuh}}, \bibinfo {author} {\bibfnamefont
  {W.}~\bibnamefont {Wegscheider}}, \bibinfo {author} {\bibfnamefont
  {K.}~\bibnamefont {Ensslin}}, \ and\ \bibinfo {author} {\bibfnamefont
  {G.}~\bibnamefont {Salis}},\ }\href {\doibase 10.1103/PhysRevB.82.235320}
  {\bibfield  {journal} {\bibinfo  {journal} {Phys. Rev. B}\ }\textbf {\bibinfo
  {volume} {82}},\ \bibinfo {pages} {235320} (\bibinfo {year}
  {2010})}\BibitemShut {NoStop}%
\bibitem [{\citenamefont {{Walser}}\ \emph
  {et~al.}(2012{\natexlab{b}})\citenamefont {{Walser}}, \citenamefont
  {{Siegenthaler}}, \citenamefont {{Lechner}}, \citenamefont {{Schuh}},
  \citenamefont {{Ganichev}}, \citenamefont {{Wegscheider}},\ and\
  \citenamefont {{Salis}}}]{Walser2012PRB}%
  \BibitemOpen
  \bibfield  {author} {\bibinfo {author} {\bibfnamefont {M.~P.}\ \bibnamefont
  {{Walser}}}, \bibinfo {author} {\bibfnamefont {U.}~\bibnamefont
  {{Siegenthaler}}}, \bibinfo {author} {\bibfnamefont {V.}~\bibnamefont
  {{Lechner}}}, \bibinfo {author} {\bibfnamefont {D.}~\bibnamefont {{Schuh}}},
  \bibinfo {author} {\bibfnamefont {S.~D.}\ \bibnamefont {{Ganichev}}},
  \bibinfo {author} {\bibfnamefont {W.}~\bibnamefont {{Wegscheider}}}, \ and\
  \bibinfo {author} {\bibfnamefont {G.}~\bibnamefont {{Salis}}},\ }\href@noop
  {} {\bibfield  {journal} {\bibinfo  {journal} {Phys. Rev. B}\ }\textbf
  {\bibinfo {volume} {\textbf{86}}},\ \bibinfo {pages} {195309} (\bibinfo
  {year} {2012}{\natexlab{b}})}\BibitemShut {NoStop}%
\bibitem [{\citenamefont {Poshakinskiy}\ and\ \citenamefont
  {Tarasenko}(2015)}]{Poshakinskiy2015}%
  \BibitemOpen
  \bibfield  {author} {\bibinfo {author} {\bibfnamefont {A.~V.}\ \bibnamefont
  {Poshakinskiy}}\ and\ \bibinfo {author} {\bibfnamefont {S.~A.}\ \bibnamefont
  {Tarasenko}},\ }\href {\doibase 10.1103/PhysRevB.92.045308} {\bibfield
  {journal} {\bibinfo  {journal} {Phys. Rev. B}\ }\textbf {\bibinfo {volume}
  {92}},\ \bibinfo {pages} {045308} (\bibinfo {year} {2015})}\BibitemShut
  {NoStop}%
\bibitem [{\citenamefont {{Mal'shukov}}\ and\ \citenamefont
  {{Chao}}(2000)}]{Malshukov2000}%
  \BibitemOpen
  \bibfield  {author} {\bibinfo {author} {\bibfnamefont {A.~G.}\ \bibnamefont
  {{Mal'shukov}}}\ and\ \bibinfo {author} {\bibfnamefont {K.~A.}\ \bibnamefont
  {{Chao}}},\ }\href@noop {} {\bibfield  {journal} {\bibinfo  {journal} {Phys.
  Rev. B}\ }\textbf {\bibinfo {volume} {\textbf{61}}},\ \bibinfo {pages}
  {R2413(R)} (\bibinfo {year} {2000})}\BibitemShut {NoStop}%
\bibitem [{\citenamefont {Kiselev}\ and\ \citenamefont
  {Kim}(2000)}]{Kiselev2000}%
  \BibitemOpen
  \bibfield  {author} {\bibinfo {author} {\bibfnamefont {A.~A.}\ \bibnamefont
  {Kiselev}}\ and\ \bibinfo {author} {\bibfnamefont {K.~W.}\ \bibnamefont
  {Kim}},\ }\href {\doibase 10.1103/PhysRevB.61.13115} {\bibfield  {journal}
  {\bibinfo  {journal} {Phys. Rev. B}\ }\textbf {\bibinfo {volume} {61}},\
  \bibinfo {pages} {13115} (\bibinfo {year} {2000})}\BibitemShut {NoStop}%
\bibitem [{\citenamefont {Moriya}\ \emph {et~al.}(2014)\citenamefont {Moriya},
  \citenamefont {Sawano}, \citenamefont {Hoshi}, \citenamefont {Masubuchi},
  \citenamefont {Shiraki}, \citenamefont {Wild}, \citenamefont {Neumann},
  \citenamefont {Abstreiter}, \citenamefont {Bougeard}, \citenamefont {Koga},\
  and\ \citenamefont {Machida}}]{Moriya2014}%
  \BibitemOpen
  \bibfield  {author} {\bibinfo {author} {\bibfnamefont {R.}~\bibnamefont
  {Moriya}}, \bibinfo {author} {\bibfnamefont {K.}~\bibnamefont {Sawano}},
  \bibinfo {author} {\bibfnamefont {Y.}~\bibnamefont {Hoshi}}, \bibinfo
  {author} {\bibfnamefont {S.}~\bibnamefont {Masubuchi}}, \bibinfo {author}
  {\bibfnamefont {Y.}~\bibnamefont {Shiraki}}, \bibinfo {author} {\bibfnamefont
  {A.}~\bibnamefont {Wild}}, \bibinfo {author} {\bibfnamefont {C.}~\bibnamefont
  {Neumann}}, \bibinfo {author} {\bibfnamefont {G.}~\bibnamefont {Abstreiter}},
  \bibinfo {author} {\bibfnamefont {D.}~\bibnamefont {Bougeard}}, \bibinfo
  {author} {\bibfnamefont {T.}~\bibnamefont {Koga}}, \ and\ \bibinfo {author}
  {\bibfnamefont {T.}~\bibnamefont {Machida}},\ }\href {\doibase
  10.1103/PhysRevLett.113.086601} {\bibfield  {journal} {\bibinfo  {journal}
  {Phys. Rev. Lett.}\ }\textbf {\bibinfo {volume} {113}},\ \bibinfo {pages}
  {086601} (\bibinfo {year} {2014})}\BibitemShut {NoStop}%
\bibitem [{\citenamefont {{Dyakonov}}\ and\ \citenamefont
  {{Perel'}}(1972)}]{Dyakonov1972}%
  \BibitemOpen
  \bibfield  {author} {\bibinfo {author} {\bibfnamefont {M.~I.}\ \bibnamefont
  {{Dyakonov}}}\ and\ \bibinfo {author} {\bibfnamefont {V.~I.}\ \bibnamefont
  {{Perel'}}},\ }\href@noop {} {\bibfield  {journal} {\bibinfo  {journal} {Sov.
  Phys. Solid State}\ }\textbf {\bibinfo {volume} {\textbf{13}}},\ \bibinfo
  {pages} {3023} (\bibinfo {year} {1972})}\BibitemShut {NoStop}%
\bibitem [{\citenamefont {Winkler}\ \emph {et~al.}(2002)\citenamefont
  {Winkler}, \citenamefont {Noh}, \citenamefont {Tutuc},\ and\ \citenamefont
  {Shayegan}}]{Winkler2002}%
  \BibitemOpen
  \bibfield  {author} {\bibinfo {author} {\bibfnamefont {R.}~\bibnamefont
  {Winkler}}, \bibinfo {author} {\bibfnamefont {H.}~\bibnamefont {Noh}},
  \bibinfo {author} {\bibfnamefont {E.}~\bibnamefont {Tutuc}}, \ and\ \bibinfo
  {author} {\bibfnamefont {M.}~\bibnamefont {Shayegan}},\ }\href {\doibase
  10.1103/PhysRevB.65.155303} {\bibfield  {journal} {\bibinfo  {journal} {Phys.
  Rev. B}\ }\textbf {\bibinfo {volume} {65}},\ \bibinfo {pages} {155303}
  (\bibinfo {year} {2002})}\BibitemShut {NoStop}%
\bibitem [{\citenamefont {Minkov}\ \emph {et~al.}(2005)\citenamefont {Minkov},
  \citenamefont {Sherstobitov}, \citenamefont {Germanenko}, \citenamefont
  {Rut}, \citenamefont {Larionova},\ and\ \citenamefont
  {Zvonkov}}]{Minkov2005}%
  \BibitemOpen
  \bibfield  {author} {\bibinfo {author} {\bibfnamefont {G.~M.}\ \bibnamefont
  {Minkov}}, \bibinfo {author} {\bibfnamefont {A.~A.}\ \bibnamefont
  {Sherstobitov}}, \bibinfo {author} {\bibfnamefont {A.~V.}\ \bibnamefont
  {Germanenko}}, \bibinfo {author} {\bibfnamefont {O.~E.}\ \bibnamefont {Rut}},
  \bibinfo {author} {\bibfnamefont {V.~A.}\ \bibnamefont {Larionova}}, \ and\
  \bibinfo {author} {\bibfnamefont {B.~N.}\ \bibnamefont {Zvonkov}},\ }\href
  {\doibase 10.1103/PhysRevB.71.165312} {\bibfield  {journal} {\bibinfo
  {journal} {Phys. Rev. B}\ }\textbf {\bibinfo {volume} {71}},\ \bibinfo
  {pages} {165312} (\bibinfo {year} {2005})}\BibitemShut {NoStop}%
\bibitem [{\citenamefont {Park}\ \emph {et~al.}(2013)\citenamefont {Park},
  \citenamefont {Shin}, \citenamefont {Song}, \citenamefont {Chang},
  \citenamefont {Han}, \citenamefont {Choi},\ and\ \citenamefont
  {Koo}}]{Park2013}%
  \BibitemOpen
  \bibfield  {author} {\bibinfo {author} {\bibfnamefont {Y.~H.}\ \bibnamefont
  {Park}}, \bibinfo {author} {\bibfnamefont {S.-H.}\ \bibnamefont {Shin}},
  \bibinfo {author} {\bibfnamefont {J.~D.}\ \bibnamefont {Song}}, \bibinfo
  {author} {\bibfnamefont {J.}~\bibnamefont {Chang}}, \bibinfo {author}
  {\bibfnamefont {S.~H.}\ \bibnamefont {Han}}, \bibinfo {author} {\bibfnamefont
  {H.-J.}\ \bibnamefont {Choi}}, \ and\ \bibinfo {author} {\bibfnamefont
  {H.~C.}\ \bibnamefont {Koo}},\ }\href {\doibase
  http://dx.doi.org/10.1016/j.sse.2013.01.016} {\bibfield  {journal} {\bibinfo
  {journal} {Solid-State Electronics}\ }\textbf {\bibinfo {volume} {82}},\
  \bibinfo {pages} {34 } (\bibinfo {year} {2013})}\BibitemShut {NoStop}%
\bibitem [{\citenamefont {Nakamura}\ \emph {et~al.}(2012)\citenamefont
  {Nakamura}, \citenamefont {Koga},\ and\ \citenamefont
  {Kimura}}]{Nakamura2012}%
  \BibitemOpen
  \bibfield  {author} {\bibinfo {author} {\bibfnamefont {H.}~\bibnamefont
  {Nakamura}}, \bibinfo {author} {\bibfnamefont {T.}~\bibnamefont {Koga}}, \
  and\ \bibinfo {author} {\bibfnamefont {T.}~\bibnamefont {Kimura}},\ }\href
  {\doibase 10.1103/PhysRevLett.108.206601} {\bibfield  {journal} {\bibinfo
  {journal} {Phys. Rev. Lett.}\ }\textbf {\bibinfo {volume} {108}},\ \bibinfo
  {pages} {206601} (\bibinfo {year} {2012})}\BibitemShut {NoStop}%
\end{thebibliography}
\end{document}